# COWPEA (Candidates Optimally Weighted in Proportional Election using Approval voting)

Toby Pereira

## Abstract

This paper describes COWPEA (Candidates Optimally Weighted in Proportional Election using Approval voting), a method of proportional representation that uses approval voting. It has also been used in other contexts and been referred to as Random Priority, though it is underdeveloped in the literature. COWPEA optimally elects an unlimited number of candidates with potentially different weights to a body, rather than giving a fixed number equal weight. A non-deterministic Approval-Based Committee (ABC) version that elects a fixed a number of candidates with equal weight is known as COWPEA Lottery. This is the only proportional ABC method known to pass the criteria of monotonicity, strong candidate Pareto efficiency, multi-winner participation, Independence of Irrelevant Ballots (IIB), Independence of Universally Approved Candidates (IUAC), Independence of Irrelevant Alternatives (IIA) and independence of clones. It is also possible to convert COWPEA and COWPEA Lottery to use score ballots. They are compared against Optimal PAV and Optimal PAV Lottery, which have good overall criterion compliance but fail monotonicity. This paper introduces Perfect Representation In the Limit (PRIL) as a new proportionality criterion.

## 1. Introduction

Finding the "Holy Grail" proportional method that can use approval or score ballots involves finding a method that is proportional according to reasonable proportionality criteria, as well as behaving properly in other ways, such as responding positively to extra votes given to a candidate and obeying certain symmetries. There was no a priori guarantee that a method fitting all relevant criteria would be possible.

The main initial goal is to find an approval-based committee (ABC) method that fulfils the desired criteria. An ABC method is one that uses approval ballots to elect a fixed number of candidates with equal weight. Once this is done, it can be extended to a more general cardinal method, where differing scores are given to candidates (e.g. from 0 to 10), rather than a simple approval or not. This paper discusses score conversions in section 7.

This paper also considers optimal, or portioning, methods. An optimal method is one that can elect any number of candidates, each with any proportion of the committee weight. Each ABC method will converge upon an optimal method in the limit as the number of elected candidates increases, in the case where unlimited candidate clones exist, or where votes are for parties fielding unlimited candidates.

Optimal methods are also interesting in this context as they can be trivially converted into non-deterministic ABC methods by using the proportions as probabilities. Also an optimal method is not simply an election method but an attempt to answer the question: how do we determine the mathematically optimal candidate weights in a proportional election? There is existing literature already on portioning methods, so this paper takes a look at this. If there is an ABC method that meets the Holy Grail criteria, then the corresponding optimal method will likely have already been described.

# 2. Criteria

There are many criteria that one might demand for an approval-based method of proportional representation. There are, however, eight main criteria that will be considered in this paper for an ABC method. They are Perfect Representation In the Limit (PRIL), monotonicity, strong candidate Pareto efficiency, the multi-winner participation criterion, Independence of Irrelevant Ballots (IIB), Independence of Universally (or Unanimously) Approved Candidates (IUAC), Independence of Irrelevent Alternatives (IIA) and independence of clones. For an optimal method, IUAC does not apply since any universally approved candidates should take all the weight between them. These criteria will now be discussed in more detail.

## 2.1 Perfect Representation In the Limit (PRIL)

There is a proliferation of proportionality criteria that have been defined in an attempt to capture the essence of proportionality (see e.g. Lackner & Skowron, 2023). However, most of these require a method to satisfy lower quota. This would rule out the Sainte-Laguë party list method (and equivalently the Webster apportionment method), and by extension any approval method that reduces to it under party-style voting, which is a problem that Lackner and Skowron acknowledge:

> As the Sainte-Laguë method is in certain aspects superior to the D’Hondt method [...] it would be desirable to have notions of proportionality that are agnostic to the underlying apportionment method. (p. 107)

The Sainte-Laguë method is considered by many to be the most mathematically proportional method (see e.g. Balinski & Peyton Young, 2010; Benoit, 2000), so to use criteria that disqualify it would be to throw the baby out with the bathwater.

Sometimes a method can pass certain strict, deterministic proportionality criteria, but fail if a sequential approximation of the method is used for computational reasons. To then disqualify the sequential method from being proportional would arguably be an overreaction, considering that it would be moving in the right direction as more candidates are elected, and might only marginally fail only fail due to an approximation. It could also mean finding in favour of an overall inferior method because its approximation happens to hit a certain “checkpoint”.

The common thread among proportionality criteria is the notion that a faction that comprises a particular proportion of the electorate should have approved that same proportion of the elected body. Indeed, taken to its logical conclusions, each voter individually can be seen as a faction of $1/v$ of the electorate for $v$ voters, and so in the ideal case it should be possible to divide the elected committee equally among the voters (where a single candidate can be divided among voters).

The Perfect Representation criterion (Sánchez-Fernández et al., 2017) essentially says that for $v$ voters, if there exists a set of candidates that would allow each voter to be able to be uniquely assigned to $1/v$ of the representation, approved by them, then such a set must be the winning set.

Voters and candidates are both treated as divisible in the assignment process. $1/v$ of the representation will be less than a whole candidate in most elections. See their paper for the formal definition.

In some ways this can be seen as a weak criterion as it only kicks into action if a committee that gives Perfect Representation exists, which might not be very often in real-life cases. However, there are cases where it can be seen as too strong, as can be seen by the following example:

**Example 1**:

2 to elect

99 voters: *AB* (*A* and *B* are the two candidates approved by these voters)
99 voters: *AC*
1 voter: *B*
1 voter: *C*

Perfect Representation demands the election of *BC*. This is despite *A* having 99% of the support, with *B* and *C* each having just 50%. The two factions of 99 voters could be increased in size arbitrarily with Perfect Representation still demanding the same result.

It makes sense that if the size of the committee is the same as the size of the electorate (or indeed a multiple of it), then a deterministic method should provide Perfect Representation, as long as it possible from the ballot profile (each voter has approved enough candidates). At that point there is never any need for compromise, as each voter can essentially get a candidate each. However, for a much smaller committee it does not intuitively seem right to make this demand whenever it is possible; in this example there are not two factions of 50% of the electorate demanding each of *B* and *C* to be elected.

Instead of Perfection Representation itself, this paper uses the Perfect Representation In the Limit (PRIL) criterion. Because it makes demands in the limit, rather than at any specific point, it is also suitable for non-deterministic methods.

PRIL: As the number of elected candidates increases, then for *v* voters, in the limit it should be possible to uniquely assign each voter to $1/v$ of the candidates, approved by them, as long as it is possible from the ballot profile. Therefore:

$1/k\ (\mathrm{var}_{i=1,\ldots,v}\ (n_i)) \rightarrow 0$ as $k \rightarrow \infty$ where $n_i$ is the number of candidates assigned to voter *i* and *k* is the number of elected candidates.

The $1/k$ term is necessary because, for example, if there are just two parties with half the votes each alternately being awarded seats, the absolute variance would just alternate between two fixed values (one of these being 0), rather than approach 0 over time.

If an ABC method passes PRIL, then the equivalent optimal method automatically will, since it is equivalent to the case where the number of elected candidates is increased in the limit (and with unlimited candidate clones allowed). For an optimal method, there is no difference between Perfect Representation and PRIL. As such, the “In the Limit” part becomes redundant, but for consistency this paper will use the term PRIL generally.

PRIL should be seen as a fairly uncontroversial criterion, in that it is not too demanding, and no method calling itself proportional should fail it, which is why it is being used as the proportionality criterion in this paper. A downside is that it does not define anything about the route to Perfect Representation, only that it must be reached in the limit as the number of candidates increases, which makes it weak in that regard. However, in that respect it has similarities with independence of clones, which is a well-established criterion. Candidates are only considered clones of each other if they are approved on exactly the same ballots (or ranked consecutively for ranked-ballot methods). We would also want a method passing independence of clones to behave in a sensible manner with near clones, but it is generally trusted that unless a passing method has been heavily contrived then it would do this. Similarly, one would expect the route to Perfect Representation in a method passing PRIL to be smooth and sensible unless a method is heavily contrived, and it would generally be fairly clear when this is the case. A heavily contrived method would also likely fail one or more of the other criteria specified in this paper.

Regardless, it is clear that the criteria demanding that lower quota be satisfied are not fit for purpose, so it is necessary to start afresh in any case. PRIL also demands only a direction of travel rather than a specific result at a specific point, which also makes it more suitable for non-deterministic methods. It may be possible to build upon PRIL and make more rigorous demands without throwing anything important away, but that is beyond the scope of this paper. Although to reiterate, if the size of the committee is the same as the size of the electorate, or a multiple of it, then a deterministic method should provide Perfect Representation, as long as it possible from the ballot profile. Otherwise it is at best a semi-proportional method.

## 2.2 Monotonicity

A method is monotonic if adding an approval for a candidate while leaving everything else unchanged does not harm the candidate’s chances of being elected. Conversely, removing an approval should not help their chances of being elected. This is a fairly simple criterion, though fairly weak, which brings us to:

## 2.3 Strong candidate Pareto efficiency

Candidate *A* Pareto dominates candidate *B* if *A* is approved on all the ballots that *B* is and at least one other ballot. Passing strong candidate Pareto efficiency means that an election method must not elect a Pareto dominated candidate unless the candidate that dominates it is also elected. For an optimal method, the Pareto dominated candidate must not be elected with any weight.

This criterion is being used as a proxy for a stronger form of monotonicity where adding an approval for a candidate should actively count in a candidate’s favour rather than merely not count against. In the case where a candidate is cloned and the clone is then approved on one extra ballot, it demands that the clone is favoured over the original candidate. Monotonicity only demands that an extra approval for a candidate does not harm the candidate.

Perhaps counterintuitively, it is actually possible to pass strong candidate Pareto efficiency and fail monotonicity (as can be seen in example 6 in section 4.1), so having both together is stronger than having either alone.

**Example 2**:

2 to elect

99 voters: *AB*
99 voters: *AC*

This example is the same as example 1, except without the two voters who did not approve *A*. It is possible for a monotonic method (and also one passing weak candidate Pareto efficiency) to be indifferent between any of the possible winning sets: *AB*, *AC*, *BC*. A method passing strong candidate Pareto efficiency must elect a set containing the universally approved candidate *A*. Methods that are indifferent between these results are immediately vulnerable to the 2 extra voters that would turn example 2 into example 1. A method that positively prefers the results containing candidate *A* would have some protection against these extra votes.

In a similar manner to PRIL, a method could be contrived so that each extra approval gives such a tiny improvement to the candidate set that it would not ever make a difference to the result of an election, except effectively as a tie-break. But as with the PRIL case, such a contrived method would likely be obvious and would also likely causes failures to other criteria. Therefore, passing strong candidate Pareto efficiency should protect against the *BC* result in example 1. A non-deterministic method that passes this criterion could still elect *BC* in example 1 (but not example 2), albeit with low probability.

## 2.4 Multi-winner participation

An extra ballot being cast in an election should not cause the number of elected candidates approved on that ballot to go down, or for a lottery method reduce the probability of at least $k$ candidates (approved on that ballot) being elected for any value of $k$. This means that a voter should never be able to make the result worse for themselves by participating in the vote.

## 2.5 Independence of Irrelevant Ballots (IIB)

A ballot approving all or none of the candidates (an irrelevant ballot) should not make a difference to the outcome or probability profile. More generally, a ballot that does not distinguish between two specific candidates should not cause any gains by either candidate at the other’s expense.

For symmetry reasons, it makes sense that a voter that is indifferent between candidates should not have any influence in distinguishing between them in the elected committee. Furthermore, as seen in section 7, passing this criterion is important in the conversion from approval-based methods to score-based methods.

For single-winner approval voting, a common rebuttal to the claim that it violates one-person-one-vote is that approving more candidates does not mean more voting power, and specifically that approving all the candidates is the same as approving none of them. For a proportional approval method, passing IIB keeps this guarantee.

Janson (2018) refers to IIB a special type of consistency.

## 2.6 Independence of Universally Approved Candidates (IUAC)

The addition and election of one or more universally approved candidates along with the addition of the right number of extra seats for them to fill should not make any difference to the probability of each of the other candidates getting elected.

IUAC can be seen as a sort of orthogonal version of IIB, and is another form of symmetry. Under IIB, voters who have approved every candidate are essentially ignored. Under IUAC, candidates who have been approved by every voter are ignored when allocating the remaining seats. Passing IUAC means that only the variable preferences are taken into account during the proportional allocation. Its importance can be seen with the following non-election example:

Three people share a house. Two prefer apples and one prefers oranges. One of the apple-preferrers does the shopping and buys three pieces of fruit. But instead of buying two apples and an orange, they buy three apples. The shopper reasons that the larger faction should have twice as much as the smaller faction when everything is taken into account, not just items where there are preferences. Because all housemates have tap water available to them, the shopper factored this into the proportional calculations. Taken to its logical conclusions, this thinking could potentially mean always awarding the largest faction everything because there is so much that we all share – air, water, public areas etc. IUAC seems a fairer demand for an allocation method.

## 2.7 Independence of Irrelevant Alternatives (IIA)

If a new candidate is added to the election and this candidate is not elected (or receives zero weight for an optimal method), and the voters' preferences regarding the other candidates remain unchanged, then the overall result should not change.

## 2.8 Independence of clones

If a set of *x* candidates is approved on exactly the same ballots, we will call them a "clone set" of size *x*. For an ABC method, the probability of each value from 0 to *x*-1 candidates from the clone set being elected must not change if the clone set is then increased in size. The probability of at least *x* members being elected also must not change. For an optimal method, the overall share of the clone set in the elected body must remain constant for any non-zero members of the clone set.

In addition to this, increasing or decreasing the number of clones in a clone set should not cause candidate *A* to be elected over candidate *B*, where neither *A* nor *B* are in the clone set.

Essentially, cloning candidates should not be to any faction's advantage or disadvantage, except where a faction does not already have enough candidates standing for their proportional share, meaning that cloning their candidates can help them out in this case.

## 2.9 Excluded criteria

Two criteria that are specifically not included are the strong multi-winner Pareto efficiency criterion and consistency.

**Strong multi-winner Pareto efficiency**: A set of candidates *S* Pareto dominates set *S′* if every voter has approved at least as many candidates in *S* as *S′* and least one voter has approved more in *S*. For a method to pass the criterion, *S′* must not be the elected set. See Lackner and Skowron (2023, p. 35) for a formal definition.

The multi-winner version of the Pareto efficiency criterion is an extension of candidate Pareto efficiency. The multi-winner version works on the assumption that a voter's satisfaction with a result is entirely dependent on the number of elected candidates that they have approved. However, the validity of this assumption is not clear.

**Example 3**:

2 to elect

1 voter: *AB*
49 voters: *BC*
50 voters: *CD*

Under the above assumption, the results *AC* and *BD* would be equivalent. Under each result, each voter would have approved exactly one of the two elected candidates. However, under the *AC* result, 1 voter has approved *A* and 99 have approved *C*. Under *BD*, it is a 50/50 split. The *AC* result is highly disproportional. It is also at least arguable that a voter would prefer to be in the position where they have an elected candidate all to themselves. By adding in a single extra voter for *A* only, a multi-winner Pareto efficient method must prefer the highly disproportional *AC* to the more balanced *BD*.

The criterion is only relevant in cases where a voter strictly prefers the case where they get a greater proportion of the representation, and this is not entirely clear for elections to public office. See also example 8 in section 5.3.

Given this, perhaps it could be argued that the multi-winner participation criterion is not important for the same reason, as they both work on the assumption that more elected representatives is strictly better. However, it is much harder to imagine a case where a participation violation is justified than it is for multi-winner Pareto. There does not appear to be anything analogous to examples 3 or 8 for participation, for example.

**Consistency**: If two separate elections give the same result (or probability profile in a lottery method), then combining the ballots for a single election should give the same result (or probability profile).

While this may seem a reasonable criterion, it can prove restrictive, and see examples 9 to 11 in section 5.3 to show its debatable utility.

# 3. Existing ABC Methods

Thorvald N. Thiele's Proportional Approval Voting (PAV) (Thiele, 1895) is the only existing method that will be discussed in detail in this paper. Other well-known methods include Phragmén's Voting Rules (Phragmén, 1899), Fully Proportional Representation (also known as Monroe's Method, Monroe, 1995), and Chamberlin-Courant Rule (first described by Thiele, ibid., and later by Chamberlin and Courant, 1983). However, these methods all lack resoluteness and therefore strong criterion compliance. The non-sequential versions would all elect *BC* in example 1, and in example 2 where *A* is universally approved, they would be indifferent between any of the three possible winning sets: *AB*, *AC*, *BC*. These methods can be modified to give them a greater responsiveness to approvals, but such add-ons tend to break other symmetries and do not give all the required criterion compliances. See also Janson (2018) for an English language discussion of PAV and Phragmén's Voting Rules.

There are also subtractive quota methods, which elect candidates sequentially and remove the quota of votes required to elect each candidate (e.g. a Hare or Droop quota). However, while some of these methods may be of practical interest, they are less so from a theoretical and criterion-compliance point of view, given their crude method of operation and lack of optimisation of any particular measure. This is why they are not being considered in this paper.

## 3.1 Proportional Approval Voting (PAV)

Proportional Approval Voting gives satisfaction scores to voters based on the number of elected candidates that they have approved. If a voter has approved *n* elected candidates, their satisfaction score is $1 + \frac{1}{2} + \ldots + \frac{1}{n}$ (this is the harmonic series). The set of candidates that gives the highest sum of satisfaction scores is the winning set. It is a generalisation of the D'Hondt party list method. There are variants with different divisors (such as one which generalises the Sainte-Laguë party list method), but these do not affect the compliance for the criteria discussed in this paper.

This deterministic ABC version of PAV passes monotonicity, strong candidate Pareto efficiency, multi-winner participation, IIB, IIA and independence of clones, but fails both PRIL and IUAC. Since each extra approval for a candidate simply adds to the sum of satisfaction scores of all the sets containing that candidate, it passes monotonicity and strong candidate Pareto efficiency fairly trivially.

PAV passes the multi-winner participation criterion. PAV works by adding up the satisfaction scores for each voter under each potential candidate set and electing the set with the highest total. Adding an extra ballot would result in a greater increase in the score of sets that include more candidates approved on this ballot. It is impossible, therefore, for participation to be violated.

IIB compliance was demonstrated by Janson (2018, see pp. 44-46).

PAV passes IIA fairly trivially since the score for set of candidates is determined independently of candidates not in that set.

PAV passes independence of clones. Since the score for set of candidates is determined independently of candidates not in that set, the number of clones in a clone set will not change the number of these clones that PAV will preferentially elect, or affect other candidates.

Example 4 shows a failure of IUAC:

**Example 4**:

6 to elect

2 voters: *U1-U3; A1-A3*
1 voter: *U1-U3*; *B1-B3*

(The *U* is for universally approved candidates in this example. However, example 5 takes a more relaxed approach to the *U* marker, using it for candidates that are approved by more than one faction, but not necessarily all.)

Without loss of generality, assuming numerical priority for lower numbers, with 3 seats and no *U* candidates, PAV would proportionally elect *A1-A2*, *B1*. To pass IUAC, therefore, PAV should elect *U1-U3*, *A1-A2*, *B1*. However, it prefers *U1-U3*, *A1-A3* over the IUAC-compliant result by a score of 6.73 to 6.65 (2 × harmonic (6) + 1 × harmonic (3) versus 2 × harmonic (5) + 1 × harmonic (4)). At an intuitive level, this is because PAV gives the result where the 2 voters have double the number of elected candidates of the lone voter (6 to 3). The IUAC-compliant result gives them 5 and 4 candidates respectively.

PAV's IUAC failure is exactly of the nature described in non-election example from section 2.6 involving fruit and tap water.

In addition to this, using a modified version of example 4, PAV can even be shown to fail PRIL.

**Example 5**:

20 to elect

2 voters: *U1-U10; A1-A10*
2 voters: *U1-U10; B1-B10*
1 voter: *C1-C20*

In this case, proportionally speaking, the *UA* and *UB* factions should have 16 elected candidates between them, and the *C* voter should have 4 candidates. The *U* candidates clearly give a higher satisfaction score than either the *A* or *B* candidates because of the Pareto dominance, so would be elected in preference to them. The winning candidate set should therefore be *U1-U10*, *A1-A3*, *B1-B3*, *C1-C4*. However, under PAV, *U1-U10*, *A1-A2*, *B1-B2*, *C1-C6* gives a higher satisfaction score (4 × harmonic (12) + 1 × harmonic (6) = 14.86, versus 4 × harmonic (13) + 1 × harmonic (4) = 14.80). Under this result, the 4 members of the *UA* and *UB* factions get only 14 candidates, whereas the lone *C* voter gets 6.

Again, at an intuitive level, this is because the PAV-preferred result of *U1-U10*, *A1-A2*, *B1-B2*, *C1-C6* gives the *UA* voters 12 elected candidates, *UB* voters also 12, and the *C* voter 6, which fits in with the faction sizes, not taking into account the overlap. The proportional result of *U1-U10*, *A1-A3*, *B1-B3*, *C1-C4* awards 13, 13, 4, which looks disproportional to PAV. This result is not because a lack of seats has caused a rounding error. This is essentially PAV's version of proportionality. Being in overlapping factions counts against voters under PAV.

This example could be extended in size indefinitely, which would show a PRIL failure (i.e. in the limit), as opposed to merely a proportionality failure in this limited case. You would simply multiply the number of candidates to elect and the number of each type of candidate by a constant. There are also 20 candidates to be elected, which is an exact multiple of the 5 voters. As PAV is a deterministic method and it is possible to achieve from the ballot profile, Perfect Representation should be obeyed in this case. Because of this, PAV is perhaps best described as a semi-proportional method.

While excluded from the Holy Grail criteria, it is perhaps worth mentioning multi-winner Pareto efficiency and consistency, both of which PAV passes.

Since a voter's satisfaction score is based purely on the number of elected candidates they have approved, any Pareto dominated set will have a lower score than the set that dominates it.

The PAV score for a potential result would simply be the sum of the two scores for this result in the two sub-elections. The highest such score in one sub-election plus the highest score in the other must also be the highest possible combined score. Therefore if each sub-election returns the same winning candidate set, this candidate set must win in the combined election, demonstrating consistency.

# 4. Optimal Methods

As previously discussed, optimal, or portioning, methods allow any number of candidates to be elected with any weight.

## 4.1 Optimal PAV

Under Optimal PAV, the candidate proportions can be found by finding the proportions in the limit as the number of elected candidates increases, under the assumption that candidates can be cloned without limit.

Optimal PAV is equivalent to the Nash Product Rule, except in cases where any voters have approved zero candidates (see Peters, 2025). The Nash Product Rule works by maximising the product of individual voters' representation (number of elected candidates they have approved).

Intuitively this result makes sense because maximising the product of the voters' representation levels is equivalent to maximising the sum of the logs of these, and the harmonic function of $x$ (which PAV uses) converges to $\ln x$ + the Euler–Mascheroni constant (approximately 0.577), and the constant proportionally disappears as you increase the number of seats.

However, if a voter has approved zero candidates, then they will be represented by zero seats even as the total number of seats gets large so any product will always be zero. In such a case, it would make sense to exclude such voters when applying the Nash Product Rule, and this would restore the equivalence.

Under Optimal PAV, the PRIL failure from example 5 is eliminated fairly trivially as the Pareto dominated *A* and *B* candidates are completely subsumed by *U*. This then gives us two neat factions – *U* and *C*. Under any strongly Pareto efficient (candidate or multi-winner version) optimal method, the IUAC criterion is not applicable since the universally approved candidates would take all the weight in the committee.

Not all election examples are so trivial, but it can be shown that Optimal PAV is proportional in general, therefore passing PRIL (see e.g. Bogomolnaia et al., 2005, which discusses the Nash Product Rule specifically).

Optimal PAV retains PAV's other passed criteria in a fairly trivial manner, except that, perhaps surprisingly, it fails monotonicity (see Brandl et al., 2021). This is shown in the following example in their paper:

**Example 6**:

1 voter: *A*
1 voter: *AB*
1 voter: *AC*
2 voters: *BCD*
2 voters: *BD*
2 voters: *CD*

In this example, Optimal PAV awards *A* 33.3% and *D* 66.7%, with *B* and *C* taking no share. However, if the *A*-only voter adds an approval for *D*, the ratios become *A*: 17.1%; *B*: 8.6%; *C*: 8.6%; *D*: 65.8%. This means that the extra approval for *D* has reduced their overall share.

The possibility of this non-monotonic result can be seen at an intuitive level. Consider just the two winning results above. We can call the original result *X*, and the result that comes from the extra *D* approval *Y*. Going by representation level, the single variable voter prefers the *X* result regardless of whether they approve *D*. The voter's representation levels under each result are as follows:

*A* only: *X*: 33.3%; *Y*: 17.1%; difference = 16.2%; ratio = 1.95:1
*A* and *D*: *X*: 100%; *Y*: 82.9%; difference = 17.1%; ratio = 1.21:1

As can be seen above, the arithmetic difference between the representation for this voter under the *X* and *Y* results is greater in the case where they approve both *A* and *D*, suggesting an even greater preference for the original result *X* after the extra approval of *D*. However, there is a greater ratio in the case where they approve just *A*. Since the Nash Product Rule maximises the product, rather than the sum, of voters' representation levels, it is the ratio that determines the difference in satisfaction score under Optimal PAV. Therefore under the assumptions of PAV and the Nash Product Rule, this voter prefers *X* to *Y* by a greater amount in the case where they only approve *A*. Therefore by approving both *A* and *D*, it tips the scales towards the *Y* result when other voters' preferences are also taken into consideration.

This monotonicity failure does not contradict strong candidate Pareto efficiency. Imagine instead of the *A*-only voter adding an approval for *D*, no changes are made to the approvals of existing candidates. Instead, a new candidate, *D'* is added. Every voter who approves *D* also approves *D'*, and the single *A*-only voter also approves *D'*. In this case, the original *D* candidate would get none of the share. *D'* would take the 65.8% originally taken by *D* with the modified ballots. Therefore strong candidate Pareto efficiency is preserved.

It is also worth noting that elections are primarily to serve voters, not candidates, so arguably the multi-winner participation criterion (which Optimal PAV passes) takes primacy over monotonicity.

## 4.2 Optimal PAV Lottery

Optimal PAV can also be converted into Optimal PAV Lottery, a sequential ABC method. Each candidate's Optimal PAV weight is used as the probability of being the next candidate elected. The probabilities are recalculated after each candidate is elected. This method is non-deterministic, but it passes all of PRIL, strong candidate Pareto efficiency, multi-winner participation, IIB, IIA and independence of clones, inheriting these trivially from Optimal PAV. It also passes IUAC, as any universally approved candidates would be elected to the first positions, and the election would proceed from there as if those candidates had never existed. However, like Optimal PAV, it is not monotonic.

It is worth now also revisiting example 1 with the PAV methods.

**Example 1**:

2 to elect

99 voters: *AB*
99 voters: *AC*
1 voter: *B*
1 voter: *C*

Deterministic PAV would elect *AB* in this example. Optimal PAV would elect *A* with 0.98 of the weight, with *B* and *C* getting 0.01 each. Under Optimal PAV Lottery, *A* would be elected alongside either *B* or *C* in 4999 out of 5000 elections. This is in stark contrast to methods that would elect *BC* despite the near-universal support for *A*.

## 4.3 PAV with party voting

If PAV is used for party-list elections, where each party fields as many candidates as there are seats available, it passes PRIL, as it is just a granular approximation of Optimal PAV, which it would reach in the limit. Despite the monotonicity failure, no other known deterministic method comes close to its criterion compliance overall in this setting. It is worth reiterating that PAV passes strong candidate Pareto efficiency, unlike some monotonic methods. If one were to devise a measure of overall positive responsiveness to votes, as opposed to looking for single points of failure, PAV with party voting (along with Optimal PAV and its lottery method) would likely strongly outperform methods electing *BC* in example 1, regardless of whether these other methods were technically monotonic. It also passes multi-winner participation, which is arguably the more important criterion of the two.

Therefore, for an approval-based party-list election where a deterministic method is required, PAV is arguably the only game in town.

## 4.4 Other optimal methods

In general, any optimal method can be turned into a non-deterministic ABC method that passes the same criteria fairly simply, by using the weights as probabilities. However, optimal methods do not in general have a neat deterministic conversion the way that Optimal PAV converts to normal PAV.

Other optimal methods that could be converted into non-deterministic ABC methods include (see e.g. Bogomolnaia et al., 2005; Brandl et al. 2021; Aziz et al., 2024):

The Conditional Utilitarian Rule is where a voter is assigned to the candidate that has the most approvals altogether of those that they approved. This assignment is equally split if there is a tie. Candidates are then weighted by the number of voters assigned to them.

The Uncoordinated Equal Shares Rule involves equally splitting a voter's assignment among all candidates that they approve and weighting candidates accordingly.

The Maximum Payment Rule sequentially assigns voters to the most approved candidate among the remaining voters if voters are eliminated from the count once they have been assigned to a candidate. Candidates are weighted accordingly.

Alternatively voters can be assigned to candidates in a way that maximises the per voter average group size (the number of voters assigned to each candidate). This is equivalent to maximising the sum of the squares of the group sizes. Candidates are then weighted by the number of voters assigned to them.

Without going into great detail on each of these methods, among other individual failures, all of them fail IIB.

**Example 7**:

2 voters: *A*
1 voter: *B*
6 voters: *AB*

A proportional method passing IIB would elect *A*:*B* in a 2:1 ratio. The above methods lack this level of sophistication in their algorithms, and none would elect in this ratio. The Uncoordinated Equal Shares Rule would give a 5:4 ratio, and the others would all give an 8:1 ratio. Under the assumption that more representation is strictly better (which would apply in many cases, but that this paper has questioned under election scenarios), it is quite arguable that the Nash Product Rule is uniquely the best portioning rule.

However, there is one optimal method known to fulfil this paper's criteria, including monotonicity.

# 5. COWPEA and COWPEA Lottery

## 5.1 COWPEA

COWPEA stands for "Candidates Optimally Weighted in Proportional Election using Approval voting" and it is an optimal method. The proportion of the weight each candidate gets in the elected body is equal to their probability of being elected in the following lottery:

Start with a list of all candidates. Pick a ballot at random and remove from the list all candidates not approved on this ballot. Pick another ballot at random, and continue with this process until one candidate is left. Elect this candidate. If the number of candidates ever goes from >1 to 0 in one go, ignore that ballot and continue. If any tie cannot be broken, then elect the tied candidates with equal probability.

This is similar to a tie-break mechanism for single-winner score voting proposed by Smith & Smith (2007):

> Choose a ballot at random, and use those ratings to break the tie. (I.e. if the tied candidates are A and B, and the randomly chosen ballot scores A higher than B, then A wins.) In the unlikely event this ballot *still* indicates that some or all of the tied candidates are tied, then one chooses at random again, and continues until the number of tied candidates is reduced to a unique winner.

This method is also referred to as Random Priority. Bogomolnaia et al. (2005) and Aziz et al. (2019) discuss it in the context of collective choice more generally, rather than for elections to public office. The context is relevant given that the multi-winner Pareto efficiency criterion is of assumed importance in these papers, and so its failure of this criterion is counted against it.

It is also mentioned by Brill et al. (2022) in conjunction with e.g. the D'Hondt party-list method in the context of party voting, which is also not the context of this paper, and it is only given brief attention.

Having brought into question the importance of the multi-winner Pareto efficiency criterion, and given the other properties of this method, along with its lack of attention previously, it is perhaps worth a revival and rebrand. Hence COWPEA.

Because each voter's ballot would be the starting ballot on $1/v$ of occasions (for $v$ voters), it would always be possible for each voter to be uniquely assigned to $1/v$ of the representation, approved by them, (as long as voters have all approved at least one candidate), meaning that it passes PRIL. Furthermore, if the ballot at the start of an iteration of the algorithm has approved more than one candidate, then the candidates approved on that ballot would be weighted in this same proportional manner according to the rest of the electorate, and so on, in a recursive manner.

COWPEA is monotonic. For it to fail monotonicity, there would have to be a possible iteration of the lottery where candidate $A$ gets elected and where adding $A$ to a particular ballot would then prevent this election of $A$. In such a case, this could only happen when this particular ballot is picked at some point in the random process. For this ballot (in the case where it does not approve $A$) to get picked and for $A$ to still get elected, none of the other uneliminated candidates could also be approved on that ballot, so that the process can continue where $A$ is eventually elected. Now imagine that $A$ is approved on the ballot. In this scenario, $A$ would get elected as soon as this ballot is picked, as none of the other remaining candidates are approved on this ballot. So a non-monotonic case is impossible and approving a candidate can never decrease their weight in the elected body.

It also passes strong candidate Pareto efficiency. If a candidate is Pareto dominated by another candidate then the weight of the dominated candidate would be zero.

COWPEA passes the multi-winner participation criterion. For it to fail, a ballot's presence in the lottery process would have to actively prevent a candidate approved on the ballot from getting elected, on least some iterations of the lottery. For the ballot to cause any effect at all, it could only happen when this particular ballot is picked at some point in the random process. If, when this ballot is picked, one or more of the candidates approved on this ballot still remains in the process, then one of these candidates is guaranteed to be elected. The ballot would then have at worst a neutral effect (one of these candidates may have been elected in this iteration anyway). If, however, when this ballot is picked, none of the candidates approved on this ballot still remains in the process, then the ballot will be ignored and it will have no effect, positive or negative. Therefore, it is impossible for a ballot to reduce the overall probability of one of its approved candidates to be elected in an iteration of the lottery and for a negative participation event to happen. Therefore COWPEA passes the multi-winner participation criterion.

COWPEA passes IIB. If a ballot has approved none or all of the remaining candidates at some stage of the lottery, then it would change nothing before the next ballot is looked at. It would effectively be ignored. Indeed, for any two candidates still in the process, a ballot that approves both or neither will not favour one over the other if it is picked in the lottery process.

COWPEA passes IIA. Only Pareto dominated candidates would be elected with no weight, so would be irrelevant in this sense. These candidates would have no effect on which other candidates are eliminated from the lottery process and could not affect the result.

As with any optimal method, IUAC does not apply to COWPEA since any universally approved candidates would take all the weight in the elected body between them. All other candidates would be Pareto dominated and would have no weight. COWPEA therefore passes the paper’s criteria for optimal methods.

## 5.2 COWPEA Lottery

COWPEA Lottery is the non-deterministic ABC conversion of COWPEA. COWPEA Lottery is simply the method that runs the lottery *k* times for *k* candidates to be elected.

As with Optimal PAV Lottery, COWPEA Lottery is non-deterministic and cannot guarantee proportionality in a given election, but it passes the PRIL criterion.

COWPEA Lottery is also monotonic, passes strong candidate Pareto efficiency, multi-winner participation IIB, IIA and independence of clones, inheriting these from COWPEA. It also passes IUAC. If there are any universally approved candidates, then they would be elected to the first positions. The election would then continue to run the same as if there had been no universally approved candidates, as with Optimal PAV Lottery. COWPEA Lottery therefore passes all the paper’s criteria for ABC methods. This is the only method known to do so. COWPEA Lottery also has the computational advantage over most other methods, including Optimal PAV Lottery, that the full probabilities do not need to be calculated in order to run an election.

## 5.3 Examples

As with PAV, it is also worth noting how COWPEA and COWPEA Lottery treat example 1:

**Example 1**:

2 to elect

99 voters: *AB*
99 voters: *AC*
1 voter: *B*
1 voter: *C*

COWPEA would elect *A* with 0.9801 of the weight, with the remaining 0.0199 shared equally between *B* and *C* (198/200 × 99/100 = 0.9801). COWPEA Lottery would elect *BC* with a probability of 0.000199, which is less than 1 in 5000 ((198/200×1/100+2/200) × 1/100 = 0.000199). *AB* and *AC* would share the rest of the probability equally. *A* therefore has more than a 4999 in 5000 probability of being elected. These numbers are very similar to those for Optimal PAV and Optimal PAV Lottery, with *A* slightly advantaged under the COWPEA methods.

As mentioned, COWPEA and COWPEA Lottery fail the multi-winner Pareto efficiency criterion and it is perhaps worth looking at an example of this.

**Example 8**:

100 voters: *AC*
100 voters: *AD*
100 voters: *BC*
100 voters: *BD*
1 voter: *C*
1 voter: *D*

COWPEA would elect each of the candidates in roughly equal proportions (with *C* and *D* getting slightly more), and with two to elect, COWPEA Lottery could elect any pair of candidates. No candidate is Pareto dominated by any other individually, but as a pair, *CD* dominates *AB* in this multi-winner sense. According to the multi-winner version of Pareto efficiency, *C* and *D* should be elected with half the weight each, and this is what Optimal PAV would do. Optimal PAV Lottery would be guaranteed to elect only *C* and/or *D* candidates if there were unlimited clones.

This example can be seen as a 2-dimensional voting space with *A* and *B* at opposite ends of one axis, and *C* and *D* at opposite ends of the other. No voter has approved both *A* and *B* or both *C* and *D*. Viewed like this, electing only *C* and *D* seems restrictive and arguably does not make best use of the voting space, which may include policy areas not considered by all candidates. It is in any case not obviously necessary to elect only *C* and *D*, even if the multi-winner Pareto criterion is useful in at least some settings.

COWPEA also fails the consistency criterion, which is passed by Optimal PAV. This failure means that combining the ballots from two elections that each produce the same result can give a different result.

**Example 9**:

2 voters: *AC*
1 voter: *A*
3 voters: *B*

**Example 10**:

3 voters: *A*
2 voters: *BC*
1 voter: *B*

In both the above examples, *C* is Pareto dominated (by *A* in example 9 and by *B* in example 10), and COWPEA would elect *A* and *B* with half the weight each in both elections (*C* would have no weight in either). We can now combine the ballots:

**Example 11**:

4 voters: *A*
4 voters: *B*
2 voters: *AC*
2 voters: *BC*

Passing the consistency criterion would commit COWPEA to electing *A* and *B* with half the weight each, as in the previous examples (which is what Optimal PAV would do). COWPEA actually elects *A* and *B* with $\frac{4}{9}$ of the weight each and *C* with $\frac{1}{9}$ of the weight, therefore failing consistency. However, *C* is no longer Pareto dominated in this example, and combining the ballots sets has therefore changed *C*'s position within the electoral landscape. It does not seem unreasonable to elect *C* with some weight in this election, and it is therefore not clear that passing the consistency criterion is strictly necessary for a proportional approval method. It is not an obvious blunder.

It is also arguably harder to justify a monotonicity failure than a failure of the multi-winner Pareto efficiency criterion or consistency in an election to public office. And given the monotonicity failures of Optimal PAV and Optimal PAV Lottery, this leaves us with just COWPEA and COWPEA Lottery as the only known methods that fulfil our requirements for optimal and ABC proportional methods respectively.

# 6. Further discussion of COWPEA and Optimal PAV

It would be easy, based on the above, to conclude in favour of COWPEA and COWPEA Lottery over Optimal PAV and Optimal PAV Lottery.

Optimal PAV does fail monotonicity, but since voting methods are there to serve voters rather than candidates and it passes participation, it is perhaps not quite the indictment it would first appear to be.

The Nash Product Rule is on a firm mathematical grounding and comes out better than Random Priority in the portioning literature under their strict assumptions, including that someone will always prefer more representation, regardless of other factors. While COWPEA may be justified to some extent in not adhering to this criterion in election settings, it comes with no mathematical measure of what should be preferred and why.

COWPEA is simply a process that produces a result. It does not give a measure of quality to all results and elect the one with the highest score. Because of this, it is much harder to answer questions of the quality of winning and non-winning results, and it appears on a weaker mathematical footing in this sense. Optimal PAV, on the other hand, can give a quality score to all results, and we can see exactly how much it prefers *CD* to *AB* and everything in between in example 8, for example. COWPEA gives the result it does and says no more. This is also why there is no known sensible way to translate COWPEA into a deterministic ABC method. Having said that, deterministic PAV is only a semi-proportional method, except for party voting, reducing the size of this point in its favour.

COWPEA has a simple algorithm and clearly passes this paper's criteria. Therefore the above criticisms could be seen as inconsequential. Running elections using COWPEA Lottery would also be far more practical than using Optimal PAV Lottery, which would likely be impossible, practically speaking, for anything other than very small elections.

# 7. Cardinal Ballots

So far we have just looked at methods that use approval ballots. However, COWPEA and Optimal PAV, along with their lottery methods, can be converted to score-based methods, for full cardinal proportional representation. Mathematically defining proportional representation to everyone's agreement is hard enough with approval voting (see section 2.1), but with score voting it is taken up another notch still. However, there are still certain things we can demand.

Any conversion should keep the same criterion compliance as the underlying approval-based method, and add scale invariance. Since scores can be seen as a proxy for utilities, only the relative difference between the scores, and not the ratios, is important. This means that if the scores on every ballot are multiplied by a positive constant the result should remain unchanged. Also if some or all voters add a constant to the score of every candidate on their ballot, the result should remain unchanged. The following ballots should therefore give the result of *A*:*B* in a 2:1 ratio (regardless of the maximum allowable score):

**Example 12**:

1 voter: *A*=6; *B*=5
1 voter: *A*=1; *B*=0
1 voter: A=9; *B*=10

This is because the score difference between *A* and *B* is constant on all ballots (at 1), with *A* being preferred by 2 ballots to 1.

If an approval-based method passes IIB, then there is a transformation that will give this form of scale invariance. This transformation converts scores into approvals, by "splitting" a voter into *s* parts (numbered from 1 to *s*) for a maximum score of *s*. Part *n* of *s* approves all candidates given a score of *n* or higher. Once this is done, an approval-based method can then be applied to the ballots.

Assuming scores out of 10, the ballots in example 12 would be converted to:

0.1 voters: *A*
0.5 voters: *AB*
0.4 voters: -

0.1 voters: *A*
0.9 voters: -

0.1 voters: *B*
0.9 voters: *AB*

Combining them together:

0.2 voters: *A*
0.1 voters: *B*
1.4 voters: *AB*
1.3 voters: -

For an IIB-compliant method, this is equivalent to:

0.2 voters: *A*
0.1 voters: *B*

This would give us the 2:1 ratio required. The maximum score does not matter here and the result would be the same regardless, as long as it is no lower than the highest score given to a candidate. However, without the guarantee of IIB, the 1.4 *AB* voters and even the 1.3 blank voters could affect the result.

Given that COWPEA and Optimal PAV, along with their lottery forms, pass IIB, the transformed versions of these methods therefore pass the form of scale invariance required. For fully optimal proportional representation with cardinal ballots, COWPEA or Optimal PAV transformed in this way are the leading contenders.

## 8. Determinism and the Holy Grail

COWPEA passes the paper's criteria for optimal methods – PRIL, monotonicity, strong candidate Pareto efficiency, multi-winner participation, IIB, IIA and independence of clones. Optimal PAV passes all except for monotonicity.

COWPEA Lottery passes the criteria for ABC methods, which includes IUAC in addition to the above. Optimal PAV Lottery fails only monotonicity.

With party-voting, deterministic PAV keeps the same criterion compliance as Optimal PAV Lottery (including PRIL), so is an ABC method that passes all but monotonicity. But this is a very limited case.

But what about a deterministic ABC method that allows for any approval ballots? Perhaps it would be possible to find a deterministic ABC method that corresponds to COWPEA in the same way that PAV corresponds to Optimal PAV. However, it is worth remembering that PAV itself does not pass the paper's proportionality criterion despite Optimal PAV doing so, and there is no guarantee that such a conversion of COWPEA would pass all of the relevant criteria either. And given that COWPEA is a process without a measure of quality, it might be impossible to reverse engineer such a method in any case.

It also seems unlikely that there could be an alternative ABC method implementation of Optimal PAV that does pass the criteria. PAV and Optimal PAV are known as welfarist methods, as their quality function is based purely on the number of elected candidates each voter has approved.

However, Peters and Skowron (2020), in showing that all welfarist methods fail laminar proportionality, give a pair of examples where all welfarist methods must give a disproportional result in at least one of them. Therefore, any deterministic ABC method implementation of Optimal PAV that passes PRIL would have to turn it into a non-welfarist method. Such a major modification would essentially turn it into a different method and would be unlikely to preserve all of its other desirable properties.

If there were to be a deterministic Holy Grail ABC method, it must also have a corresponding optimal or portioning method that it would reach in the limit. But given the absence in the portioning literature of any suitable methods beyond COWPEA and Optimal PAV, it is perhaps unlikely that there could be a third method that would bear any fruit in this endeavour, which would limit our options.

It seems likely that if a deterministic ABC method that passes the “Holy Grail” criteria were to exist, it would be fairly clean and simple, and would have been found by now. However, it still remains an open question in the absence of any proof against. And the task would still remain to prove exactly which criteria are compatible and incompatible with each other.

## 9. Discussion and Conclusions

COWPEA and COWPEA Lottery are the only methods known to fulfil the criteria set out at the start of this paper for optimal and ABC methods respectively, and they can be converted to cardinal-ballot elections relatively simply. Optimal PAV and Optimal PAV Lottery fail only monotonicity. There is no known deterministic “Holy Grail” ABC method.

While COWPEA and Optimal PAV give optimal candidate weights, COWPEA Lottery and Optimal PAV Lottery are ABC methods and could be used in elections where a fixed number of representatives with equal weight are required. They do not guarantee proportionality in an individual election, but over many elections, either through time or across many constituencies in a general election, the combined results would tend to be more proportional than that of a deterministic method.

National level proportional representation is normally achieved with party lists, but this takes some of the power away from voters, and gives less room for independent candidates. Achieving proportional representation at a national level can also be unwieldy and complex. While Optimal PAV Lottery is likely to be computationally too complex beyond very small elections, COWPEA Lottery can achieve this proportional representation in a much simpler manner than party lists are able to. It also enables people to vote for individual candidates rather than parties, and without the need to collate votes at a national level. In a case like this, non-determinism would be a feature of COWPEA Lottery, not a bug.

Calculating a result under COWPEA Lottery is computationally very simple. And while it might be considered too theoretical to be used in real-life political elections, at least for the time being, its simplicity and criterion compliance mean that it can be used right now by people wanting to run smaller elections. For example, a group of friends could use it as a way to decide on an activity, and it would ensure a certain fairness over time, without anyone having to keep track of previous decisions.

Another potential use for COWPEA Lottery is in elections where there are many elected candidates, so that any potential disproportionality caused by its random process would be minimised. Boehmer et al. (2024) discuss the performance of several ABC methods in elections on the Polkadot blockchain with committee sizes of around 300. It would be interesting to see the performance of COWPEA Lottery alongside the other methods under the same metrics.

In conclusion, COWPEA Lottery and Optimal PAV Lottery bring a new level of criterion compliance to the landscape of ABC methods. In the case of COWPEA Lottery, these criteria include monotonicity, along with added simplicity compared to Optimal PAV Lottery. The methods are of theoretical interest and, particularly in the case of COWPEA Lottery, of practical interest.

## References

Aziz, H., Bogomolnaia, A., & Moulin, H. (2019). Fair mixing: the case of dichotomous preferences. *Proceedings of the 2019 ACM Conference on Economics and Computation*, 753–781. https://doi.org/10.1145/3328526.3329552

Aziz, H., Lederer, P., Lu, X., Suzuki, M., & Vollen, J. (2024). Sequential payment rules: approximately fair budget divisions via simple spending dynamics. *arXiv:2412.02435 [cs.GT]*. https://arxiv.org/abs/2412.02435

Balinski, M. L., & H. Peyton Young. (2010). *Fair Representation: Meeting the ideal of one man, one vote*. Brookings Institution Press.

Benoit, K. (2000). Which electoral formula is the most proportional? A new look with new evidence. *Political Analysis*, *8*(4), 381–388. https://doi.org/10.1093/oxfordjournals.pan.a029822

Bogomolnaia, A., Moulin, H., & Stong, R. (2005). Collective choice under dichotomous preferences. *Journal of Economic Theory*, *122*(2), 165–184. https://doi.org/10.1016/j.jet.2004.05.005

Boehmer, N., Brill, M., Cevallos, A., Gehrlein, J., Sánchez-Fernández, L., & Schmidt-Kraepelin, U. (2024). Approval-based committee voting in practice: a case study of (over-)representation in the Polkadot blockchain. *Proceedings of the AAAI Conference on Artificial Intelligence*, *38*(9), 9519–9527. https://doi.org/10.1609/aaai.v38i9.28807

Brandl, F., Brandt, F., Peters, D., & Stricker, C. (2021). Distribution rules under dichotomous preferences: two out of three ain’t bad. *EC ’21: Proceedings of the 22nd ACM Conference on Economics and Computation*, 158-179. https://doi.org/10.1145/3465456.3467653

Brill, M., Gölz, P., Peters, D., Schmidt-Kraepelin, U., & Wilker, K. (2022). Approval-based apportionment. *Mathematical Programming, 203*(1–2), 77–105. https://doi.org/10.1007/s10107-022-01852-1

Chamberlin, J. R., & Courant, P. N. (1983). Representative deliberations and representative decisions: proportional representation and the Borda rule. *American Political Science Review*, *77*(3), 718–733. https://doi.org/10.2307/1957270

Janson, S. (2018). Phragmén's and Thiele's election methods. a*rXiv:1611.08826 [math.HO]*. https://arxiv.org/abs/1611.08826

Lackner, M., & Skowron, P. (2023). Multi-winner voting with approval preferences. In *SpringerBriefs in Intelligent Systems*. https://doi.org/10.1007/978-3-031-09016-5

Monroe, B. L. (1995). Fully proportional representation. *American Political Science Review*, *89*(4), 925–940. https://doi.org/10.2307/2082518

Peters, D. (2025). *Party Approval PAV Converges to Nash*. Retrieved 13th October, 2026, from https://dominik-peters.de/notes/party-approval-pav-converges-to-nash.pdf

Peters, D., & Skowron, P. (2020). Proportionality and the limits of welfarism. a*rXiv:1911.11747 [cs.GT]*. https://arxiv.org/abs/1911.11747

Phragmén, E. (1899). Till frågan om en proportionell valmetod. *Statsvetenskaplig Tidskrift*, *2*(2), 297–305. Retrieved 13th October, 2026, from https://www.rangevoting.org/PhragmenVoting1899.pdf

Sánchez-Fernández, L. P., Elkind, E., Lackner, M., Fernández, N., Fisteus, J. A., Val, P. B., & Skowron, P. (2017). Proportional justified representation. *Proceedings of the AAAI Conference on Artificial Intelligence*, *31*(1). https://doi.org/10.1609/aaai.v31i1.10611

Smith, T. C., & Smith, W. D. (2007, 3rd March). *Tie breaking methods*. rangevoting.org. Retrieved 13th Ocotber, 2026, from https://rangevoting.org/TieBreakIdeas.html

Thiele, T. N. (1895). Om flerfoldsvalg. *Oversigt Over Det Kongelige Danske Videnskabernes Selskabs Forhandlinger*, 415–441. Retrieved 13th October, 2026, from http://publ.royalacademy.dk/backend/web/uploads/2019-09-04/AFL%203/O_1895_00_00_1895_4465/O_1895_19_00_1895_4463.pdf